\documentclass{llncs}
\usepackage[english]{babel}
\usepackage{graphicx}
\newcommand{\newmata} [3] {\newcommand{#1}[#2]{\mbox{$#3$}}}
\newmata {\ceil}{1}{\left\lceil #1 \right\rceil} %mm

\begin{document}
\mainmatter
\title{Self-similar planar graphs as models for complex networks
\thanks{Research supported by the Ministerio de Educaci\'on y Ciencia,  Spain, and the European Regional Development Fund under project TEC2005-03575  and by the Catalan Research Council under project 2005SGR00256.  L. Chen and Z. Zhang are supported by the National Natural Science Foundation of China under  Grant No. 60704044, the Postdoctoral Science Foundation of China
under Grant No. 20060400162, and the Huawei Foundation of Science
and Technology (YJCB2007031IN).}
}
\titlerunning{Self-similar planar graphs}  % abbreviated title (for running head)

\author{Lichao Chen\inst{1}, Francesc Comellas\inst{2}\and Zhongzhi Zhang\inst{1}}
\authorrunning{L. Chen, F.  Comellas \& Z.Zhang}% abbreviated author list (for running head)
\institute{Department of Computer Science and Engineering and\\
Shanghai Key Lab of Intelligent Information Processing,\\
Fudan University, Shanghai 200433, China\\
\email{zhangzz@fudan.edu.cn}
\and
Departament de Matem\`atica Aplicada IV,\\
Universitat Polit\`ecnica de Catalunya\\
Avda. Canal Ol\'{\i}mpic s/n, 08860
Castelldefels, Catalonia, Spain
\email{comellas@ma4.upc.edu}\\
%\quad URL: \texttt{ http://www-ma4.upc.edu/\homedir comellas/
}

% ----------------------------------------------------------------
\maketitle % typeset the title of the contribution

\begin{abstract}
In this paper we introduce a family of planar, modular
and self-similar graphs which have small-world and scale-free properties.
The main parameters of this family are comparable to those of networks
associated to complex systems, and therefore the graphs are of interest % can be useful % considered
as mathematical models for these systems. % networks.
As the clustering coefficient of the graphs is zero, this family % of graphs
is an explicit construction that does not match the usual
characterization of hierarchical modular networks, % in the literature,
namely that vertices have clustering values inversely proportional
to their degrees.
\end{abstract}

%\begin{keywords}
%Complex networks,  self-similar graphs,  hierarchical graphs, modular networks.
%\end{keywords}

%%%%%%%%%%%%%%%%%%%%%%%%%%%%%%%%%%%%%%%%%%%
%%%%%%%%%%%%%%%%%%%%%%%%%%%%%%%%%%%%%%%%%%%
%%%%%%%%  1 INTRODUCTION                          %%%%%%%%%%%%%%%%%%
%%%%%%%%%%%%%%%%%%%%%%%%%%%%%%%%%%%%%%%%%%%
\section{Introduction}
Research and studies performed in the last few years show
that many networks associated with complex systems, like the
World Wide Web, the Internet, telephone networks, transportation
systems (including power and water distribution networks),  social
and biological networks, belong to a class of networks now known as small-world
scale-free networks, see~\cite{AlBa02,Ne03}  and references therein.
These networks exhibit a small  average distance and diameter
(with respect to a random network with the same number of nodes and links)
and, in many cases, a strong local clustering (nodes have many mutual neighbors).
Another important common characteristic is that the number of links attached
to the nodes usually obeys a power-law distribution (is scale-free).
By introducing a new measuring technique, it has recently been discovered
that many real networks are also self-similar, see~\cite{SoHaMa05,SoHaMa06}.
Moreover, a degree hierarchy in these networks is sometimes related to
the modularity of the system which they model.

Most of the network models considered are probabilistic, however in
recent years some deterministic models have been proposed which are very
often based on iterative constructions such that, at each step, one
or  more vertices are connected to certain subgraphs (for example,
the so called $k$-trees~\cite{BePi71}). Another technique produces graphs
by duplication of certain substructures, see~\cite{ChLuDeGa03}.
Here we propose a new family of graphs which generalize these methods
by introducing at each iteration a more complex substructure than a
single vertex. The result is a family of planar, modular, hierarchical and self-similar
graphs, with small-world scale-free characteristics and
with clustering coefficient zero.
%These graphs do  not follow the standard characterization of hierarchical graphs
%--that vertices have clustering values inversely proportional to their degrees--,
%as their clustering coefficient is zero.
We note that some important real life networks,  for example the  networks associated to
electronic circuits or Internet~\cite{Ne03}, have these characteristics as they are modular,
almost planar and with a reduced clustering coefficient and have small-world scale-free properties.
Thus, these networks  can be modeled by our construction.
A related family of graphs based on triangles, and which therefore has
a high clustering coefficient was introduced in~\cite{ZhZhFaGuZh07}.

%%%%%%%%%%%%%%%%%%%%%%%%%%%%%%%%%
%%%%%%%%%%%%%%%%%%%%%%%%%%%%%%%%%
%%%%%%%%%  2 HIERARCHICAL MODULAR GRAPHS        %
%%%%%%%%%%%%%%%%%%%%%%%%%%%%%%%%%
\section{Hierarchical modular graphs}
Several authors classify as  hierarchical graphs, graphs with a modular
structure and a strong  connectedness hierarchy of the vertices which
produces a power-law degree distribution. Moreover, they consider that
the most important signature of hierarchical modularity is given by a
clustering distribution with respect to the degree according to
$C(k) \propto 1/k$, see~\cite{BaOl04,DoGoMe02,WuRaBa03}.
In this section % we introduce
we define an analyze
a family of hierarchical modular graphs,
which are scale-free, planar and have  clustering coefficient zero.
They prove the existence of hierarchical graphs which do not have
the above-mentioned relationship between the clustering coefficient
and the degrees of the corresponding vertices.

Deterministic models for simple hierarchical networks have been
published in~\cite{No03,RaBa03}.  These models consider the recursive
union of several basic structures (in many cases, complete graphs) by adding
edges connecting them to a selected root vertex. These and other hierarchical
graphs have been considered when modeling metabolic networks
in~\cite{JeToAlOlBa00,RaSoMoOlBa02}. Hierarchical modularity also appears in
some models based on $k$-trees or clique-trees, where the graph is constructed
by adding at each step one or more vertices and each is connected independently
to a certain subgraph~\cite{DoGoMe02,CoFeRa04,ZhCoFeRo06}.
The introduction of the so-called {\em hierarchical product of graphs}
in~\cite{BaCoDaFi08} allows a generalization and a rigorous study
of some of these models.

In~\cite{SoHaMa05,SoHaMa06}, Song, Havlin and Makse relate the scale-free
and  the self-similarity properties as they verify that many self-similar
graphs associated to real life complex systems have a fractal dimension
and provide a connection between this dimension and the exponent of the
degree power-law. However, a classical scale-free mode, the preferential
attachment by Barábasi-Albert~\cite{AlBa02}, which many authors consider
a paradigm for these networks, has a null fractal dimension.
This is not a paradox as the  Barab\'asi-Albert model lacks modularity
because of its generation process based on the individual
introduction of vertices.

In the next subsection we give details of our construction which is also
based on an iterative process. However, the  introduction at each step
of a certain substructure  allows the formation of modules and results
in a final graph with a self-similar structure.

%%%%%%%%%%%%%%%%%%%%%%%%%%%%%%%%%%%%%%%%%%%%%%

\subsection{Iterative algorithm to generate the graph  $H(t)$}
The graph $H(t)$ is constructed as follows:
For $t=0$, $H(0)$ is $C_4$, a length four cycle.
We define now as {\em generating cycle} a cycle $C_4$ whose vertices
have not been introduced at the same iteration step and {\em passive cycle}
a cycle $C_4$ which  does not  verify this property.
For  $t\geq 1$, $H(t)$ is  obtained from $H(t-1)$ by considering all
their generating cycles  $C_4$ and connecting, vertex to vertex,
to each of them  a new  cycle  $C_4$.
This operation is equivalent to adding to the graph a cube $Q_3$
by identifying vertex to vertex the generating cycle with
one of the cycles of $Q_3$.
The process is repeated until the desired graph order is reached.

 %%%%%%%%%%%%%%%%%%%%%%%%%%%%%%%%%%%%%%%%%%%%%%
% Figure  HIERARCHICAL NETWORK %%%%%%%%%%%%%%%
%%%%%%%%%%%%%%%%%%%%%%%%%%%%%%%%%%%%%%%%%%%%%%
\begin{center}
\begin{figure}[htbp]
\includegraphics[width=4.2cm]{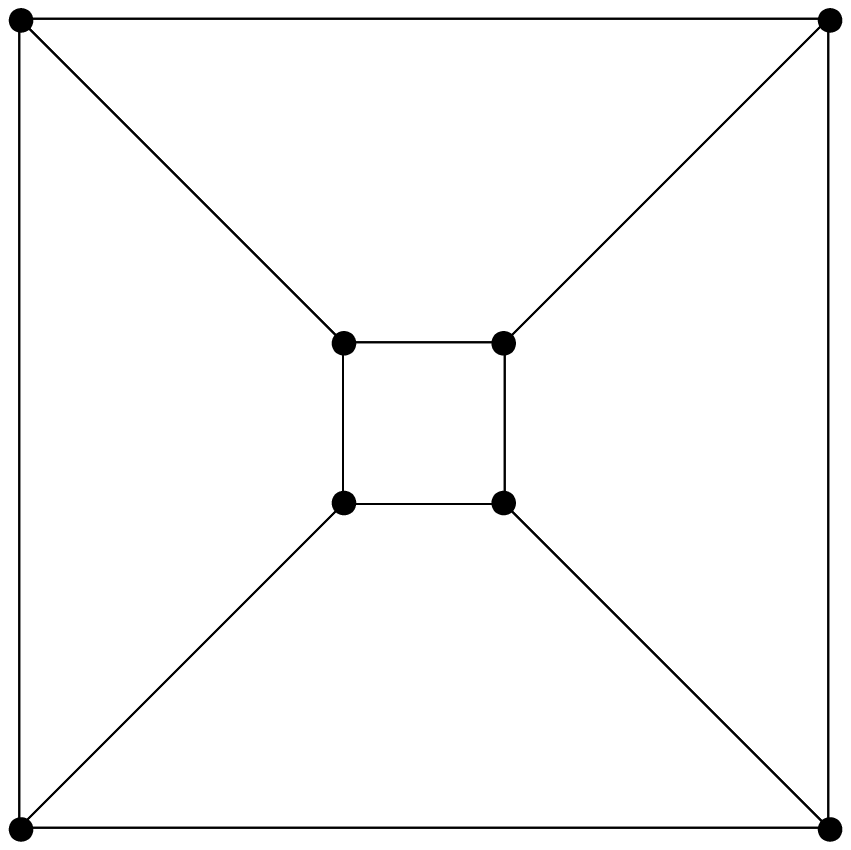}$\!\!\!\!\!$\includegraphics[width=4.2cm]{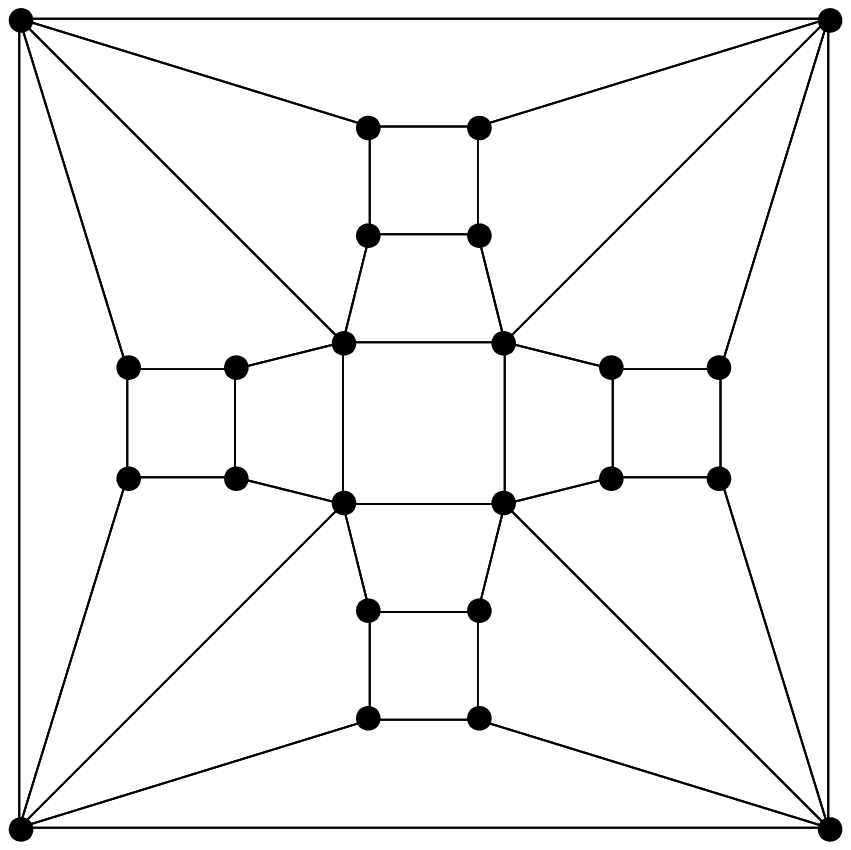}$\!\!\!\!\!$\includegraphics[width=4.2cm]{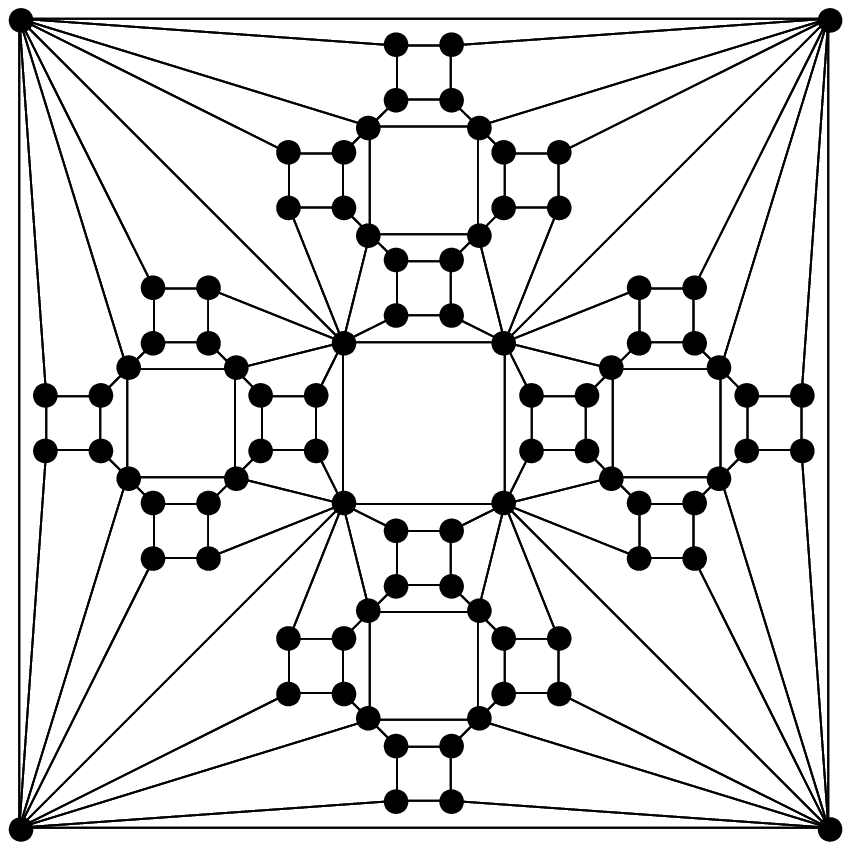}
\caption{Graphs $H(t)$ produced at  iterations $t=1, 2$ and $3$.
}
\label{fig:plana}
\end{figure}
\end{center}
%%%%%%%%%%%%%%%%%%%%%%%%%%%%%%%%%%%%%%%%%%%%%%

\subsection{Recursive modular construction}
\begin{center}
\begin{figure}[htbp]
\vskip -1cm
\includegraphics[width=3cm]{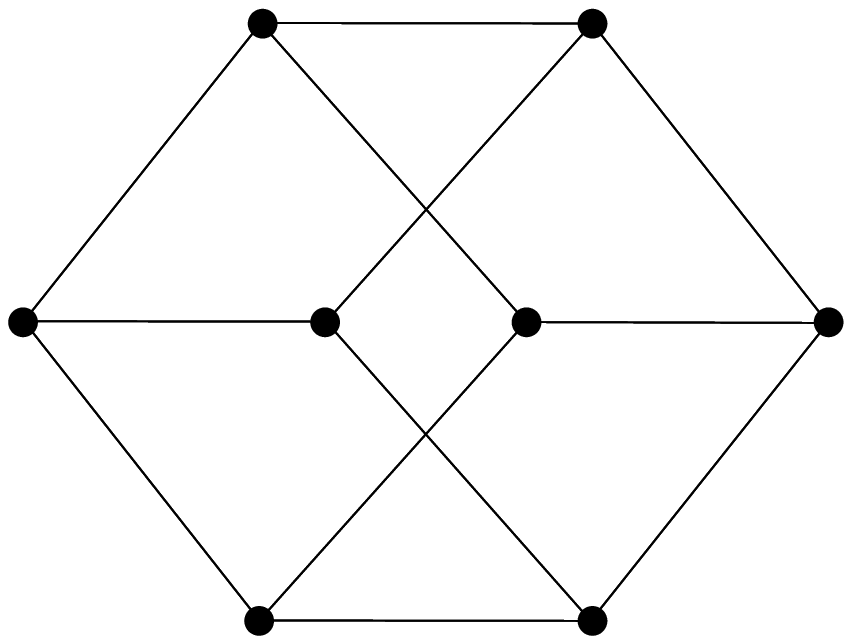} \includegraphics[width=4cm,viewport=15 15 300 270,clip]{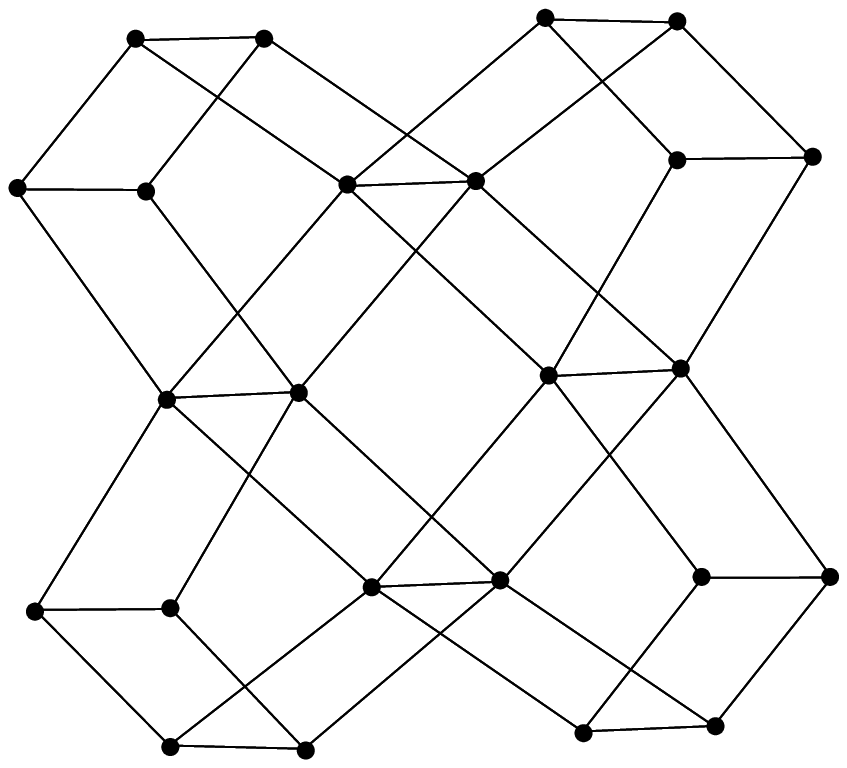}
\includegraphics[width=5.5cm,viewport=20 20 300 285,clip]{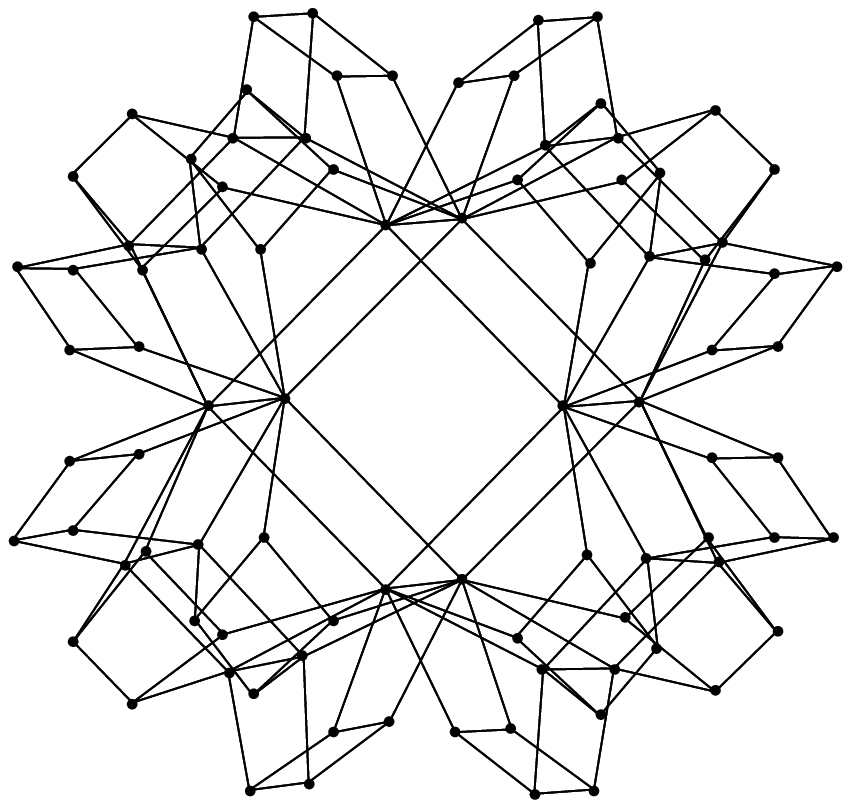}
\caption{Modular construction of $H(t)$ for $t=1,2$ y $3$.  At step  $t$,  we merge four copies of $H(t-1)$ to four cycles of the cube $Q_3$,
leaving opposite cycles free. See the text for details.
} \label{fig:recmod}
\end{figure}
\end{center}

The graph $H(t)$ can be also defined as follows:
For  $t=0$, $H(0)$ is the cycle  $C_4$.
For  $t\geq 1$,  $H(t)$  is produced from four copies of  $H(t-1)$
by identifying, vertex to vertex,  the initial passive cycle of each $H(t-1)$
with each of four  consecutive cycles of $Q_3$
(leaving two opposite cycles of $Q_3$ free),  see  Fig.~\ref{fig:recmod}.

 %%%%%%%%%%%%%%%%%%%%%%%%%%%%%%%%%%%%%%%%%%%%%%
% Figure  3 HIERARCHICAL NETWORK %%%%%%%%%%%%%%%
%%%%%%%%%%%%%%%%%%%%%%%%%%%%%%%%%%%%%%%%%%%%%%
%%%%%%%%%%%%%%%%%%%%%%%%%%%%%%%%%%%%%%%%%%%%%%

%\begin{figure}[htp]
%\centering
%\includegraphics[totalheight=0.8\textheight,viewport=50 260 400 1000,clip]{erptsqfit}
%\caption[Transverse momentum distributions - E-R model.]
%{Transverse momentum distributions - E-R model fit (intercept 1.2).}\label{fig:erptsqfit}
%\end{figure}
%
%The most useful option to use, I find, is to set the width or height you want your picture scaled to to be some fraction of the textwidth or height, as is done in the above, where we have [totalheight=0.8\textheight], that is, the total height of the figure is set to 80\,\% of the height of the text on a normal page of typing.
%
%The viewport command needs some fiddling with to get right - if you choose to use it at all (you need it for mathematica files as they are saved very very poorly as PS files). Basically it's the number of points to take off the top, left, bottom, right, in that order, I think. You just have to play around and see. The idea is, it takes this much space off all around the document, leaving only the inside bit that is what you want to see in your document. You need the ``clip'' command to force LaTeX to ignore the stuff outside the limits you specify by ``viewport''.

\subsection{Properties of  $H(t)$}
{\em Order and size of $H(t)$}.---
We use the following notation: $\tilde{V}(t)$ and $\tilde{E}(t)$ denote,
respectively,  the set of  vertices and edges introduced at the step $t$ and
$\tilde{C}(t)$  is the number of generating cycles $C_4$ at this step
(which will be used to produce the graph $H(t+1)$.

Note that at each iteration, any generating cycle is replaced by four
new generating cycles and one passive cycle.
Therefore: $\tilde{C}(t+1)=4\cdot\tilde{C}(t), t\geq1$ and $\tilde{C}(0)=1$.
Thus  $\tilde{C}(t)=4^t$.
Moreover, each generating cycle introduces at the next iteration four
new vertices and eight new edges.
As a consequence, $\tilde{V}(t)=4\cdot\tilde{C}(t-1)=4\cdot 4^{t-1}$
and $\tilde{E}(t)=8\cdot\tilde{C}(t-1)=8\cdot 4^{t-1}=2\cdot 4^t$, thus:
\begin{equation}\label{OrdSiz}
|V(t)|=\sum_{i=0}^t \tilde{V}(t)=\frac{4^{t+1}+8}{3}\nonumber \qquad
|E(t)|=\sum_{i=0}^t \tilde{E}(t)=\frac{2\cdot 4^{t+1}+4}{3}
\end{equation}
%
%%%%%%%%%%%%%%%%%%%%%%%%%%%%%%%%%%%%%%%%%%%%%%%%
%% Table ORDER, SIZE, # ACTIVE CYCLES
%%%%%%%%%%%%%%%%%%%%%%%%%%%%%%%%%%%%%%%%%%%%%%%%%
% Para tablas utillize
\begin{table}
\begin{center}
 % Para elaborar tablas con LaTeX utilice
\begin{tabular}{cccc}
\hline
Step & Vertices  & Edges &  Number of active cycles\\
\hline
& & &\\
0  & 4 & 4 & 1 \\
1  & 8 & 12 & 4  \\
2  & 24 & 44  & 16  \\
3  & 88 & 172  & 64 \\
$\cdots$ & $\cdots$ & $\cdots$ & $\cdots$ \\
t  & $\frac{4^{t+1}+8}{3}$ & $\frac{2\cdot 4^{t+1}+4}{3}$  & $4^t$ \\
$\cdots$ & $\cdots$ & $\cdots$ & $\cdots$ \\
\hline
\end{tabular}
 \end{center}
\caption{Number of vertices, edges and generating cycles of $H(t)$  at each step.}
\label{tab:ordsize}
\end{table}

{\em Degree distribution}.---
Intially, at $t=0$,  the graph is a single generating cycle $C_4$
and its four vertices have degree two.

When a new vertex $i$ is added to the graph at iteration
$t_i$ ($t_i\geq1$), it has degree $3$.
We denote by $C(i,t)$  the number of generating cycles at
iteration $t$ which will produce
new vertices that will connect to vertex  $i$ at step  $t+1$.
At iteration $t_i$, when  vertex  $i$ is introduced,
the value of  $C(i,t_i)$ is $2$.
According to the construction process of the graph, at each
iteration, each new neighbor of $i$ belongs to two generating
cycles  where  $i$  is also a vertex.
If we denote as $k(i,t)$  the degree of vertex  $i$ at step  $t$,
then we have the following relationship: $C(i,t)=k(i,t)-1$.

We now compute $C(i,t)$.
As we have seen above, each generating cycle where $i$ belongs to,
produces two new generating cycles which also have $i$ as a vertex.
Thus $C(i,t)=2\cdot C(i,t-1)$.  Using the initial condition
$C(i, t_i)=2$, we have  $C(i,t)=2^{t-t_{i}+1}$.
Therefore the degree of vertex  $i$ at the step  $t$ is
\begin{equation}\label{ki}
k(i,t)=2^{t-t_{i}+1}+1.
\end{equation}

Note that the initial four vertices of step 0 follow a different process.
In this case  $C(i,0)=2^t$ and  $k(i,t)=2^{t}+1$.
Thus,  at step  $t$ the initial four vertices of the graph have the same
degree than those introduced at step  1.

%%%%%%%%%%%%%%%%%%%%%%%%%%%%%%%%%%%%%%%%%%%%%%%%%
%% TABLA        VERTICES y GRADOS en el PASO t
%%%%%%%%%%%%%%%%%%%%%%%%%%%%%%%%%%%%%%%%%%%%%%%%%
% Para tablas utillize
%\begin{table}
%\begin{center}
% % Para elaborar tablas con LaTeX utilice
%\begin{tabular}{cccc}
%\hline
%Num. vertices  &  Degree & Step \\
%\hline\smallskip
%$4^{t}$ & 3 & $t$\\
%$4^{t-1}$ & 5 & $t-1$ \\
%$\cdots$  &  $\cdots$  & $\cdots$  \\
%$4^{t-i}$ & $2^{i+1}+1$ & $t-i$ \\
%$\cdots$  &  $\cdots$  & $\cdots$  \\
%$4$ & $2^{t}+1$ & 1 \\
%$4$ & $2^{t}+1$ & 0 \\
%\hline
%\end{tabular}
% \end{center}
% %
%\caption{ Total number of vertices for each degree and step when they were incorporated to the graph  $H(t)$.}
%\label{tab:grados}     % Emplee una sola etiqueta
%\end{table}
%%%%%%%%%%%%%%%%%%%%%%%%%%%%%%%%%%%%%%%%%%%%%%%%%
%%%%%%%%%%%%%%%%%%%%%%%%%%%%%%%%%%%%%%%%%%%%%%%%%

From equation~(\ref{ki})  we verify % and table~\ref{tab:grados}
 that
the graph has a discrete degree distribution and we use the
technique described by  Newman in ~\cite{Ne03} to find the
cumulative degree distribution $P_{\rm cum}(k)$  for a vertex with
degree $k$: $P_{\rm cum}(k)=\sum_{\tau \leq
t_i}{|V(\tau)|}/{|V(t_i)|} =({4^{t_i+1}+8)/(4^{t+1}+8})$.

Replacing $t_i$, from equation~(\ref{ki}), in the former equation
$t_i=t+1- {\ln(k-1)}/{\ln 2}$ we obtain
$P_{\rm cum}(k)=({16\cdot4^{t}\cdot(k-1)^{-2}+8})/({4^{t}+8})$,
which for large values of  $t$,  allows us to write
$P_{\rm cum}(k)\sim k^{1-\gamma_k}=  k^{-2}$,
and therefore the degree distribution, for large graphs, follows
a power-law with exponent $\gamma_{k}=3$.
Research on networks associated to electronic circuits (these
networks show planarity, modularity and a small clustering coefficient)
gives similar  values for their degree power-law
distribution~\cite{FeJaSo01,Ne03}.
More precisely, the largest  benchmark considered --a network
with 24097 nodes, 53248 edges, average degree 4.34 and average
distance 11.05--  has a degree distribution which follows a power-law
with exponent 3.0, precisely the same as in our model, and it has a small clustering
coefficient $C=0.01$.

%%%%%%%%%%%%%%%%%%%%%%%%%%%%%%%%%%%%%%%%%%%%%%%%%%%%%%%%%%
% Figure  5
%%%%%%%%%%%%%%%%%%%%%%%%%%%%%%%%%%%%%%%%%%%%%%%%%%%%%%%%%%
\begin{figure}
\begin{center}
\includegraphics[width=6cm]{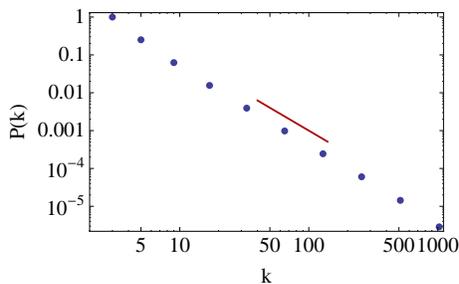}
\caption{Log-log representation of the cumulative degree distribution for $H(10)$ with
$|V|=1398104$ vertices. The reference line has slope $- 2$.} \label{fig:CumDegDist}
\end{center}
\end{figure}
%%%%%%%%%%%%%%%%%%%%%%%%%%%%%%%%%%%%%%%%%%%%%%%%%%%%%%%%%%

{\em Diameter}.---
At each step we introduce, for each generating cycle, four new
vertices which will form a new cycle $C_4$
(and these vertices are among them at maximum distance 2).
As all join the graph of the former step with one new edge, in the worst situation  the diameter will increase by exactly 2 units.
Therefore  $D_t= D_{t-1}+2$. $t\geq 2$.  As  $D_1=3$,
we have that the diameter of  $H(t)$ is $D_t=2\cdot t +1$ if  $t\geq 1$.
\smallskip

{\em Average distance}.---
The average distance of $H(t)$ is defined as:
\begin{equation}\label{apl01}
%  \bar{d}_t  =  {|V(t)|  \choose 2}^{-1}   \sum_{i,j \in V(t)} d(i,j) ={ |V(t)| \choose 2 }^{-1} S_t \,,
\bar{d}_t  =  \frac{1}{{\mbox{\scriptsize $ |V(t)| (|V(t)|-1)/2$}}}   \sum_{i,j \in V(t)} d_{i,j} \,,\end{equation}
where $d_{i,j} $ is the distance between vertices $i$ and $j$.
We will denote as  $S_t$ the sum  $ \sum_{i,j \in V(t)} d_{i,j}$.

\begin{figure}
%\vspace{-0.5cm}
\begin{center}
\includegraphics[width=11cm]{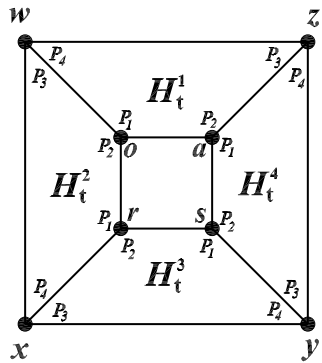}
\caption{Illustration of the classification of nodes in $H_t^{\varphi}$,  $\varphi=1,2,3, 4$.}
%\caption{Second construction method and node classification of the
%considered network. (i) $H(t+1)$ can be obtained by joining four
%copies of $H(t)$ denoted as $H_t^{\varphi}$ $(\varphi=1,2,3, 4)$,
%which are connected to one another at the edge nodes (i.e., $w$,
%$x$, $y$, $z$, $o$, $r$, $s$, and $a$). (ii) Illustration of the
%recursive definition of node classification. All interior nodes in
%$H(t)$ can be classified into four deferent parts represented as
%$P_\tau$ $(\tau=1,2,3, 4)$. From the node classification of
%$H_t^{1}$, $H_t^{2}$, $H_t^{3}$, and $H_t^{4}$, we can derive
%recursively the classification of nodes in network $H(t+1)$.}
\label{fig:copy}
\end{center}
\end{figure}

The modular recursive construction of $H(t)$ allows us to calculate the exact value of  $\bar{d}_t $.
 At step $t$, $H(t+1)$ is  obtained from the juxtaposition of four copies of
$H(t)$,  which we  label $H_t^{\varphi}$, $\varphi=1,2,3, 4$,  on top of the cube $Q_3$
(see Figs.~\ref{fig:recmod}~and~\ref{fig:copy}).
The copies are connected one to another at  the vertices   which we call
{\em connecting vertices} and we label   $w$,  $x$, $y$, $z$, $o$, $r$, $s$, and $a$. 
The other vertices of $H(t+1$ will be called {\em interior vertices}.
Thus, the sum of distances distance $S_{t+1}$ satisfies the following recursion:
\begin{equation}\label{apl03}
  S_{t+1} = 4\, S_t + \Delta_t-4.
\end{equation}
where $\Delta_t$ is the sum over all shortest paths whose endvertices are
not in the same  $H(t)$ copy  and the last term compensates
 for the overcounting of  some paths  between the connecting vertices
--for example, $d(w,o)$ is included both in  $H_t^{1}$ and $H_t^{2}$--.
Note that the paths that contribute to $\Delta_t$ must all go through at
least one of the eight connecting vertices.
The analytical expression for $\Delta_t$  is
not difficult to find.
%%%%%%%%%%%%%%%%%%%%%%%%%%%%%%%%%%%%%%%%%%%%%%%%%%%%%%%%%
% Figure  4
%%%%%%%%%%%%%%%%%%%%%%%%%%%%%%%%%%%%%%%%%%%%%%%%%%%%%%%%%%
%%%%%%%%%%%%%%%%%%%%%%%%%%%%%%%%%%%%%%%%%%%%%%%%%%%%%%%%%%
We denote as $\Delta_t^{\alpha,\beta}$  the sum of all shortest paths
with endvertices in $H_t^{\alpha}$ and $H_t^{\beta}$.
$\Delta_t^{\alpha,\beta}$ excludes the paths such that  either endvertex
is a connecting vertex.  Then the total sum $\Delta_t$ is
\begin{eqnarray}\label{apl04}
\Delta_t &=&\Delta_t^{1,2} + \Delta_t^{1,3} + \Delta_t^{1,4}+
\Delta_t^{2,3} + \Delta_t^{2,4}+ \Delta_t^{3,4}+20 \nonumber\\
&\quad&+\sum_{\stackrel{i \in H_t^{3} \cup H_t^{4},}{ i \notin x, r, s, y, a,z}} (d_{w,i}+d_{o,i})
                +\sum_{\stackrel{i \in H_t^{1} \cup H_t^{4},}{ i \notin w,o, a, z, s, y}} (d_{x,i}+d_{r,i})\nonumber\\ 
&\quad&+\sum_{\stackrel{i \in H_t^{1} \cup H_t^{2},}{ i \notin x, r, o, w, a, z}} (d_{s,i}+d_{y,i})
                +\sum_{\stackrel{i \in H_t^{2} \cup H_t^{3},}{ i \notin w, o,x, r, s, y}} (d_{a,i}+d_{z,i}),
\end{eqnarray}
where the term 20 comes from the sum of   $d_{w,s}$, $d_{w,y}$,
$d_{o,s}$, $d_{o,y}$, $d_{x,a}$, $d_{x,z}$, $d_{r,a}$, and
$d_{r,z}$.  

By symmetry, $\Delta_t^{1,2} = \Delta_t^{1,4} = \Delta_t^{2,3} =
\Delta_t^{3,4}$, $\Delta_t^{1,3}=\Delta_t^{2,4}$, and
$\sum_id_{w,i}=\sum_id_{o,i}=\sum_id_{x,i}=\sum_id_{r,i}=\sum_id_{s,i}=\sum_id_{y,i}=\sum_id_{a,i}=\sum_id_{z,i}$,
and
%\begin{equation}\label{apl05}
%\Delta_t = 4 \Delta_t^{1,2} + 2\Delta_t^{1,3}+8\,\sum_{i \in H_t^{3}
%\cup H_t^{4}\,, i \notin x, r, s, y, a, z} d_{w,i}\,.
%\end{equation}
\begin{equation}\label{apl05}
\Delta_t = 4 \Delta_t^{1,2} + 2\Delta_t^{1,3}+
8\,\sum_{\stackrel{i \in H_t^{3}\cup H_t^{4},}{ i \notin x, r, s, y, a, z} }  d_{w,i}\,.
\end{equation}

To calculate $\Delta_t$, we classify the interior vertices of
$H(t+1)$ into four different classes according to their distances
to each of the four vertices  $w$, $x$, $y$, and $z$. 
Vertices $w$, $x$, $y$, and $z$  are not classified into any 
of these classes which we represent as 
$P_{1}$, $P_{2}$, $P_{3}$, and $P_{4}$,
respectively.
This classification  is represented in Fig.~\ref{fig:copy}. 
By construction, for an arbitrary interior vertex $v$, there must exist
one of the above mentioned vertices (say $w$) satisfying $d_{v,w}<d_{v,x}$,
$d_{v,w}<d_{v,y}$, and $d_{v,w}<d_{v,z}$. 
All the interior vertices nearest to $w$ (resp. $x$, $y$, and $z$) are assigned to  class $P_{1}$
(resp. $P_{2}$, $P_{3}$, and $P_{4}$).  
The total number of vertices of  $H_{t}$ that belong to the class $P_{\tau}$ ($\tau=1,2,3,4$) 
is denoted by $N_{t,P_{\tau}}$. 
Since the four vertices $w$, $x$, $y$, and $z$  play a symmetrical  role, 
classes $P_{1}$, $P_{2}$, $P_{3}$, and $P_{4}$ are equivalent. 
Thus, $N_{t,P_1}=N_{t,P_2}=N_{t,P_3}=N_{t,P_4}$ which will be abbreviated to
 $N_{t}$ from now on. We have
\begin{equation}\label{apl06}
  N_{t} = \frac{|V_t|-4}{4}=\frac{4^t-1}{3}.
\end{equation}

We denote by $L_{t+1,P_1}$ ($L_{t+1,P_2}, L_{t+1,P_3},L_{t+1,P_4}$) the
sum of distances between vertices $w$ ($x$, $y$, $z$)
and all interior vertices $v \in P_1$ $(P_{2}, P_{3}, P_{4}) $ of  $H(t+1)$. 
Because of the symmetry,
$L_{t+1,P_1}=L_{t+1,P_2}=L_{t+1,P_3}=L_{t+1,P_4}$ that will be
written as $L_{t+1}$ for short.
%On the other hand, since the above definition of interior node classification is recursive (For instance, Class $P_{1}$ and $P_{4}$ in $H_t^{1}$, class $P_{2}$ and $P_{2}$ in $H_t^{2}$, and one shared edge node $o$ belong to Class $P_{1}$ in $H(t+1)$, see Fig.~\ref{fig:copy}.),
Taking into account the second method of  constructing $H(t)$, see Fig.~\ref{fig:copy}, we can write the
following recursive formula for $L_{t+1}$:
\begin{equation}\label{apl07}
  L_{t+1} = 4\,L_{t}+2\,N_t+1.
\end{equation}
We can solve Eq.~(\ref{apl07}) inductively, with  initial condition $L_1 = 1$,  and we have
\begin{equation}\label{apl08}
  L_{t} = \frac{1}{18} \left(3t\cdot  4^t+ 2\cdot 4^t-2\right).
\end{equation}

We now return to compute Eq.~(\ref{apl05}), with
$\Delta_t^{1,2}$   given by the sum
\begin{eqnarray}\label{apl09}
  \Delta_t^{1,2} = \!\! \sum_{\stackrel{u \in H_t^{1},u \notin \{w,o,a,z\};}{\, v \in H_t^{2},v \notin \{w,x,r,o\}}} \!\!\! d_{u,v} %\nonumber\\
   = \sum_{i=1}^{4}\sum_{j=1}^{4}
  d_{P^{t,1}_i,P^{t,2}_j}\,,
\end{eqnarray}
where $P^{t,1}_i$ and $P^{t,2}_j$ are the vertex classes $P_i$ and $P_j$
of $H_{t}^{1}$ and $H_{t}^{2}$, respectively, and
$d_{P^{t,1}_i,P^{t,2}_j}$ is the sum  of distances  $d_{u,v}$ for all
vertices  $u \in P_i \subset H_{t}^{1} $ and $v\in P_j \subset H_{t}^{2}$.

% In Eq.~(\ref{apl09}),
We have:
\begin{eqnarray}\label{apl10}
  d_{P^{t,1}_1,P^{t,2}_1} = \! \! \sum_{\stackrel{u \in P_1  \subset H_t^{1},}{\, v \in P_1  \subset H_t^{2}}}   \!\! d_{u,v}
            =\!\!  \sum_{\stackrel{u \in P_1  \subset H_t^{1},}{\, v \in P_1  \subset H_t^{2}}} \!\! ( d_{u,o}+d_{o,r}+d_{r,v})
  % &=& N_t\,\sum_{u \in P_1  \subset H_t^{1}} d_{u,o}+ N_t\,\sum_{v \in P_1  \subset H_t^{2}} d_{r,v}+ \sum_{u \in P_1  \subset H_t^{1},\, v \in P_1  \subset H_t^{2}}d_{o,r}\nonumber\\
 = 2N_t L_t+N_t^2.
\end{eqnarray}
Following the same process, we obtain $d_{P^{t,1}_i,P^{t,2}_j}$ for the  different values of  $i$ and $j$, which we use in Eq.~(\ref{apl09}) giving:
%The results are $d_{P^{t,1}_1,P^{t,2}_2}=d_{P^{t,1}_4,P^{t,2}_3}=2N_t L_t$, $d_{P^{t,1}_1,P^{t,2}_1}=d_{P^{t,1}_1,P^{t,2}_3}=d_{P^{t,1}_2,P^{t,2}_2}=d_{P^{t,1}_3,P^{t,2}_3}=d_{P^{t,1}_4,P^{t,2}_2}=d_{P^{t,1}_4,P^{t,2}_4}=2N_t L_t+(N_t)^2$, $d_{P^{t,1}_1,P^{t,2}_4}=d_{P^{t,1}_2,P^{t,2}_1}=d_{P^{t,1}_2,P^{t,2}_3}=d_{P^{t,1}_3,P^{t,2}_2}=d_{P^{t,1}_3,P^{t,2}_4}=d_{P^{t,1}_4,P^{t,2}_1}=2N_tL_t+2\,(N_t)^2$, and $d_{P^{t,1}_2,P^{t,2}_4}=d_{P^{t,1}_3,P^{t,2}_1}=2N_t L_t+3\,(N_t)^2$.
% Inserting the expressions for $d_{P^{t,1}_i,P^{t,2}_j}$ into Eq.~(\ref{apl09}), we get  $\Delta_t^{1,2} =32\,N_t  L_t+24\,N_t^2$. Similarly we obtain that  $\Delta_t^{1,3} =32\,N_tL_t+32\, N_t^2$.
\begin{equation}\label{apl11}
  \Delta_t^{1,2} =32\,N_t
L_t+24\,(N_t)^2\,.
\end{equation}
Analogously, we can obtain
\begin{equation}\label{apl12}
  \Delta_t^{1,3} =32\,N_tL_t+32\,(N_t)^2\,.
\end{equation}

Now, to find an expression for $\Delta_t$,   the only thing left is to
evaluate the last term  of  Eq.~(\ref{apl05}), which can be obtained as above
\begin{eqnarray}\label{apl13}
  8\!\!\! \sum_{\stackrel{i \in H_t^{3} \cup H_t^{4}\,, }{i \notin x, r, s, y, a, z}} \!\! d_{w,i}=64\,L_t+128\,N_t\,.%=16\,\sum_{i \in H_t^{3} \,, i \notin x, r, s, y}d_{w,i}\nonumber\\
%&=&16\,\sum_{i \in P_1\subset H_t^{3}}(d_{w,s}+d_{s,i})+16\,\sum_{i
%\in P_2\subset H_t^{3}}(d_{w,r}+d_{r,i})\nonumber\\
%&\quad& +16\,\sum_{i \in P_2\subset
%H_t^{3}}(d_{w,x}+d_{x,i})+16\,\sum_{i
%\in P_4\subset H_t^{3}}(d_{w,y}+d_{y,i})\nonumber\\
%&=&64\,L_t+128\,N_t\,.
\end{eqnarray}

%Substituting Eqs.~(\ref{apl11}),~(\ref{apl12}) and~(\ref{apl13})
%into Eqs.~(\ref{apl05}) and~(\ref{apl03}), we obtain the recursive
%expression for the total distance $S_t$:
%\begin{equation}\label{apl14}
 % S_{t+1} =4\,S_{t}+192\,N_tL_t+160\,(N_t)^2+64\,L_t+128\,N_t+16\,.
%\end{equation}
%Substituting Eq.~(\ref{apl06}) for $N_t$ and Eq.~(\ref{apl08}) for
%$L_t$ into Eq.~(\ref{apl14}), and using $S_0 = 8$, we have
%\begin{equation}\label{apl15}
 % S_{t+1} =\frac{8}{27}[10+14\times 4^t+3(t+1)\times 16^t].
%\end{equation}
%Inserting Eq.~(\ref{apl15}) into Eq.~(\ref{apl01}), the analytical
Finally, and combining the former expressions, we   write  the exact result for the average distance
of $H(t)$, $\bar{d}_t$,  as
\begin{equation}\label{apl16}
  \bar{d}_t  = \frac{4}{3}\cdot \frac{10+14\cdot 4^t+3(t+1) 16^t}{10+13\cdot 4^t+4\cdot  16^t}\,.
\end{equation}
Notice that for a large order ($t \rightarrow \infty$)
$\bar{d}_{t} \simeq t+1 \sim \ln |V_{t}|$,
%\begin{equation}\label{apl17}
%\bar{d}_{t} \simeq t+1 \sim \ln |V_{t}|,
%\end{equation}
which means that the average distance  shows a logarithmic scaling
with the order of the graph, and has a similar behavior as  the
diameter (the graph is small-world).
\smallskip

{\em Strength distribution}.---
The strength of a node in a network is associated to resources or
properties allocated to it, as the total number of publication
of an author, in the case of the network associated to the Erd\H os
number; the total number of passengers in the world-wide airports network, etc.

In our case we associate to each vertex the area of the passive cycle,
defined by the four vertices introduced at a given step.
For this purpose we assume a uniform construction of the graph.
At the initial step the area is ${\cal{A}}_0$  and we denote as ${\cal{A}}_t$
the area of the passive cycle introduced at step $t$.
By convention, we establish that the area of this cycle is one fifth
of the area of the cycle where it connects (as each introduction
of a passive cycle
is associated to the simultaneous introduction of four generating cycles).
Therefore we have ${\cal{A}}_t= (\frac{1}{5})^{t}{\cal{A}}_0$.
A vertex $i$ introduced at $t_i$ will have strength
$s(i,t_i)=(\frac{1}{5})^{t_i}{\cal{A}}_0$ and it will keep it i
n further steps $t>t_i$.
As we want to find the strength distribution for all vertices
of the graph at step $t$, we have that
$s(i,t_i)=(\frac{1}{5})^{t_i-t}\cdot{\cal{A}}_t$.

Using equation~(\ref{ki})  we obtain the following power-law for
the correlation between the strength and the degree
of a vertex:
\begin{equation}\label{s-vs-k}
s(i,t)=\frac{1}{5} {\cal{A}}_t  (k(i,t)-1)^{\ln 5/\ln 3},
\end{equation}
which for large values of the degree $k$ leads to  $s(k)\sim k^{\ln 5/\ln 3}$.

We should mention that similar exponents have been found for the relation
between the strength and the degree of the node of real life networks
like the airports network, Internet and the scientist collaboration graph~\cite{BaBaPaVe04}.

After a similar analysis to the calculation of the degree distribution,
we find that the strength distribution also follows a power law with exponent:
\begin{equation}\label{gammas01}
\gamma_{s}=1+2\frac{\ln 2}{\ln 5}.
\end{equation}
It has been shown that if a weighted graph with a non-linear correlation between
strength and degree $s(k)\sim k^\beta$   and the degree and strength distributions follows
power laws,  $P(k)\sim k^{-\gamma_{k}}$ and  $P(s)\sim s^{-\gamma_{s}}$,
then there exists a general relationship between  $\gamma_{k}$ y $\gamma_{s}$ given by
$\gamma_{s}=\frac{\gamma_{k}}{\beta}+\frac{\beta-1}{\beta}$ ~\cite{BaBaPaVe04}.

As in our case  $\gamma_{k}=3$ y $\beta={\ln 5/\ln 3}$
from the former relationship, the exponent of the strength distribution is
$\gamma_{s}=3\frac{\ln 2}{\ln 3}+\ln 2({\frac{\ln 5}{\ln 2}-1})/{\ln5}$,
and we obtain the same value $\gamma_{s}$  (\ref{gammas01}) which
was computed directly.

%%%%%%%%%%%%%%%%%%%%%%%%%%%%%%%%%%%%%
%5 CONCLUSIONS %%%%%%%%%%%%%%%%%%%%%%%
%%%%%%%%%%%%%%%%%%%%%%%%%%%%%%%%%%%%%%
\section{Conclusion}
The family of graphs introduced and studied here has as main characteristics
planarity, modularity, degree hierarchy, and small-world and scale-free properties.
At the same time the graphs have clustering zero.
A  combination  of modularity and scale-free  properties is present
in many real networks like those
associated to living organism (protein-protein interaction networks)
and some social and technical networks~\cite{RaBa03,RaSoMoOlBa02}.
The added property of a small clustering coefficient appears also in
some technological networks (electronic circuits, Internet, P2P) and
social networks~\cite{Ne03}.
Therefore our model, with a null clustering coefficient, could be
considered to model these networks and also it can be used to study
other properties without the influence of the clustering.
% isolate the influence of other properties.
The deterministic character of the family, as opposed to
usual probabilistic models, should facilitate the exact computation
of many network parameters.

On the other hand, simple variations of our model allow the introduction
of clustering.
As an example, by adding to each passive cycle an edge we can introduce
two triangles for each cycle and therefore obtain a planar graph
with non-zero clustering.
Replacing in the construction each passive cycle by a complete
graph $K_4$ will produce a family with a relatively large clustering
coefficient. However the graph will no longer be planar.

%
%\begin{acknowledgement}
% Incluya aqu\'{\i}, si los hay, los oportunos agradecimientos.
%\end{acknowledgement}

%
\end{document}